\documentclass[runningheads]{llncs}

\usepackage{subcaption,wrapfig,multirow,multicol}
\usepackage{color,tabularx, colortbl,amssymb,amsmath,bm,arydshln}
\usepackage{enumitem,kantlipsum,arydshln}
\usepackage{amssymb ,booktabs,url,tcolorbox,mathtools} 
\usepackage{newtxmath}
\usepackage{amsmath}
\usepackage{tabularx}
\usepackage{tikz}
\usepackage{ulem}
\usepackage{amssymb}
\usepackage{pifont}
\usepackage{orcidlink}
\usepackage{hyperref}
\usepackage{academicons}
\usepackage{xcolor}

\begin{document}

\title{ReFormeR: Learning and Applying Explicit Query Reformulation Patterns}

\author{%
Amin Bigdeli\inst{1}\orcidlink{0009-0003-8977-9312} \and
Mert Incesu\inst{2}\orcidlink{0009-0007-1129-2597} \and
Negar Arabzadeh\inst{3}\orcidlink{0000-0002-4411-7089} \and
Charles L. A. Clarke\inst{1}\orcidlink{0000-0001-8178-9194} \and
Ebrahim Bagheri\inst{2}\orcidlink{0000-0002-5148-6237}
}

\authorrunning{A. Bigdeli et al.}

\institute{%
University of Waterloo, Waterloo, Ontario, Canada\\
\email{abigdeli@uwaterloo.ca}, \email{claclark@gmail.com}
\and
University of Toronto, Toronto, Ontario, Canada\\
\email{mert.incesu03@gmail.com}, \email{ebrahim.bagheri@utoronto.ca}
\and
University of California, Berkeley, Berkeley, California, USA\\
\email{negara@berkeley.edu}
}

\maketitle

\begin{abstract}

We present \texttt{ReFormeR}, a pattern-guided approach for query reformulation. Instead of prompting a language model to generate reformulations of a query directly, \texttt{ReFormeR} first elicits short reformulation patterns from pairs of initial queries and empirically stronger reformulations, consolidates them into a compact library of transferable reformulation patterns, and then selects an appropriate reformulation pattern for a new query given its retrieval context. The selected pattern constrains query reformulation to controlled operations such as sense disambiguation, vocabulary grounding, or discriminative facet addition, to name a few. As such, our proposed approach makes the reformulation policy explicit through these reformulation patterns, guiding the LLM towards targeted and effective query reformulations. Our extensive experiments on TREC DL 2019, DL 2020, and DL Hard show consistent improvements over classical feedback methods and recent LLM-based query reformulation and expansion approaches. 

\end{abstract}

\section{Introduction}

Recently, Large Language Models (LLMs) have been adopted to generate reformulations and expansions of search queries \cite{wang2023query2doc,dhole2024genqrensemble,seo2025qa,lai2024adacqr,wen2024elaborative,lai2025adarewriter,ran2025two}. Early neural methods produced document side expansions and synthetic queries such as doc2query \cite{nogueira2019doc2query,gospodinov2023doc2query,nogueira2019document}, followed by approaches that synthesize query side textual surrogates such as \texttt{query2doc} \cite{wang2023query2doc,zhang2024exploring,yu2022generate,yao2024pure,kassaie2025exploiting}. These methods use generative capacity to bridge vocabulary gaps and often improve first stage recall or reranking quality \cite{jagerman2023query,zhang2024exploring}. Yet they largely treat the generator as a black box that produces useful text without articulating what transformation is being applied or why a particular reformulation to the query may be helpful.

Building on this foundation, a second wave of work aimed to add structure and control to LLM-based reformulation \cite{dhole2024genqrensemble,wang2023generative}. Generative Query Reformulation approaches use prompt-guided editing to improve effectiveness over classical baselines, with ensemble variants such as \texttt{GenQREnsemble} increasing coverage and stability through diverse prompts and sampling strategies \cite{dhole2024genqrensemble}. Methods that integrate semantic generation with retrieval, such as \texttt{MUGI}, couple dense retrieval with generated augmentations to improve matching beyond lexical overlap \cite{zhang2024exploring}. Question answer-aware expansion, such as \texttt{QA-Expand}, further conditions generation on answer cues to inject discriminative terms \cite{seo2025qa}. Despite these advances, generation often remains weakly constrained and lacks an explicit account of which transformation should be applied for a given query and context. This gap makes it difficult to guarantee stability across queries and to transfer successful reformulations on queries across collections.

In this paper, we move beyond text generation as an end in itself by placing \textit{reformulation patterns} at the center of the query reformulation process. The premise is that reformulations are effective when they realign the query’s distribution with that of the relevant set and reduce ambiguity; a system should therefore learn the patterns of useful reformulations that achieve this alignment and select among them based on observable evidence. Rather than prompting an LLM to produce query reformulations directly, we first elicit concise \textit{reformulation patterns} from pairs of initial queries and empirically stronger reformulations, consolidate these \textit{patterns} into a small library of reusable reformulation patterns, and then choose the most appropriate \textit{pattern} for a new query in light of retrieval context. Only after a \textit{pattern} is selected do we generate the reformulation under that \textit{pattern}'s constraints. This design preserves the strengths of LLM generation while making the reformulation policy explicit, and transferable across queries and collections by consistently applying the patterns.

Reformulation patterns offer two concrete benefits. They constrain query reformulations to patterns that lead to meaningful revisions to the query, such as `sense disambiguation', `controlled vocabulary grounding', or `discriminative facet addition', which may reduce query drift and improve interpretability. It also enables targeted use of patterns to justify why a particular reformulation should be applied, improving stability relative to prompt-only methods. In broad terms, our proposed approach induces a library of reformulation patterns extracted based on past successful reformulations, selects a pattern for an unseen query based on its context from the pattern library, and generates a controlled query reformulation based on that pattern. 

The major contributions of our work can be enumerated as follows: (1) A pattern-guided framework for query reformulation that induces and reuses a compact library of reformulation patterns derived from past successful query reformulations. (2) A context-aware mechanism that selects an appropriate \textit{pattern} for reformulation before generation, improving stability and reducing query drift. (3) A controlled reformulation procedure that operationalizes selected patterns and produces reformulated queries. (4) A comprehensive empirical study that situates the method against classical feedback baselines and recent LLM-based approaches, demonstrating consistent gains over various datasets.

\section{Proposed Approach}

\noindent \textbf{Rationale for Proposed Approach. } Query reformulation is effective when reformulations move the query or its representation toward that of the relevant documents. Classical information retrieval explains this effect through models that reward alignment between query terms and the language of the relevant set, including probabilistic relevance modeling, BM25 scoring, and relevance models that estimate a term distribution for relevant items. Recent LLM-based approaches achieve this and improve effectiveness, yet they often generate query reformulations without specifying the transformation being applied or why it should help for a given query and context, which limits interpretability and can induce drift.

Our proposed \texttt{ReFormeR} approach addresses this gap by inducing a compact set of reusable reformulation patterns from pairs of initial queries and empirically stronger reformulations. Each pattern captures a common reformulation approach such as ``sense disambiguation'', ``controlled vocabulary grounding'', or ``discriminative facet addition'', among others. At inference time, the system uses the retrieval context to select an appropriate pattern from the reformulation pattern library and then generates a pattern-guided reformulation. By making the reformulation policy explicit and reusable rather than implicit in prompts or term scoring heuristics, \texttt{ReFormeR} provides stable, interpretable, and context-aware reformulations.

\noindent \textbf{Approach Formalization. }\texttt{ReFormeR} learns from pairs of queries and empirically stronger reformulations and turns these observations into a compact library of reusable reformulation patterns that encode common semantic edits such as replacing a colloquial variant with a controlled vocabulary term, adding a discriminative facet to resolve sense ambiguity, or expanding an acronym. Let $\mathcal{C}$ be the document collection and let $R$ be a base retriever with scoring function $S(q,d)$ that ranks $d \in \mathcal{C}$ for a query $q$. For any $q$, denote by $D_k(q)$ the top $k$ documents returned by $R$. Let $\mathcal{Q}=\{(q_i,\tilde q_i)\}_{i=1}^{N}$ be training pairs where $\tilde q_i$ is a higher performing reformulation of $q_i$ on a disjoint development split. The objective is to induce a finite set $\mathcal{P}=\{p_1,\dots,p_M\}$ of patterns that capture these recurrent reformulations. We obtain an induction mapping $g$ that assigns each pair to one pattern, $g(q_i,\tilde q_i)\in\mathcal{P}$, by eliciting a concise description of the reformulation with an LLM for each pair and clustering the resulting representations to produce non-overlapping groups. Writing $y_i=g(q_i,\tilde q_i)$ for the assigned label, we then train a context-aware selector $s_\theta$ that, given a new query $q$ and its retrieval context $D_k(q)$, produces a distribution over patterns $\pi_\theta(p \mid q, D_k(q)) \in \Delta^{M-1}$. The selector uses the induced labels for supervision with loss as follows:
\[
\mathcal{L}_{\text{sel}}(\theta) = -\frac{1}{N}\sum_{i=1}^{N} \log \pi_\theta\!\big(y_i \mid q_i, D_k(q_i)\big).
\]
To generate a concrete reformulation, a conditional generator $G_\phi$ takes the query, its context, and a chosen pattern and returns a reformulated variant:
\[
r = G_\phi\big(q, D_k(q), p\big).
\]
The final retrieval query is the hybrid $q^{\star}= q \oplus r$, where $\oplus$ denotes concatenation that preserves the original phrasing while injecting the pattern-guided reformulation. Ranking proceeds with $S(q^{\star},d)$ for $d \in \mathcal{C}$. When graded relevance is available, let $\mathrm{Eff}(q)$ denote an effectiveness measure such as nDCG@10 for the ranking induced by $S(q,\cdot)$. Training seeks parameters that improve expected effectiveness under the selected pattern and the generated reformulation:
\[
\max_{\theta,\phi}\ \mathbb{E}_{q}\ \mathrm{Eff}\!\big(q^{\star}\big)
\quad\text{with}\quad 
p \sim \pi_\theta(\cdot \mid q, D_k(q)),\ 
r = G_\phi(q, D_k(q), p),\ 
q^{\star}= q \oplus r.
\]
In practice, we optimize $\mathcal{L}_{\text{sel}}$ together with a supervised sequence loss for $G_\phi$ using $(q_i, D_k(q_i), y_i, \tilde q_i)$, which ties the rationale for reformulation to the concrete reformulation and yields a single process that explains \textit{why} a reformulation should help and \textit{how} it is applied.

\section{Experiments and Results}

All code, data, and prompts are publicly available on our GitHub Repository\footnote{\url{https://github.com/aminbigdeli/ReFormeR}}.

\begin{figure}[t]

\begin{tcolorbox}[
    colback=gray!4!white,
    colframe=gray!70!black,
    sharp corners,
    boxrule=0.4pt,
    width=\linewidth,
    boxsep=1pt,
    left=4pt,
    right=4pt,
    bottom=2pt,
    top=2pt,
    colbacktitle=gray!60!white,
    coltitle=black,
    title=\textbf{Prompt for Extracting Reformulation Patterns},
    fonttitle=\bfseries
]
\scriptsize 

\textbf{System:} You are QueryReformulationLLM, an intelligent assistant that identifies and updates abstract patterns that describe how queries are reformulated to improve retrieval effectiveness.

\smallskip
\textbf{User:} Given a set of query reformulation pairs below and optional prior list of consolidated patterns, your objectives are: \begin{enumerate}[leftmargin=*,itemsep=1pt,topsep=1pt] \item Identify the transformation pattern(s) underlying each reformulation. \item Consolidate the global pattern set by merging semantically similar strategies and refining their names and descriptions. \end{enumerate}
\smallskip

Query Reformulation Pairs: \{query\_pairs\}

Consolidated Patterns: \{existing\_patterns\} 
\smallskip

Each extracted pattern should include a pattern name, an informative description, a generalized transformation rule, and representative examples.  
Return the results of consolidated patterns:
\smallskip

\{"Consolidated Patterns": [...]\}

\end{tcolorbox}

\caption{Overview of prompt used to extract/consolidate reformulation patterns.}
\label{fig:reformulation_prompt}

\end{figure}

\subsection{Experimental Setup}
\noindent \textbf{Datasets and Metrics.} Our experiments evaluate the proposed framework on three benchmark datasets, namely TREC DL 2019~\cite{TREC2019}, TREC DL 2020~\cite{TREC2020}, and TREC DL Hard~\cite{DL_Hard}. These datasets encompass queries with comprehensive graded relevance judgments designed to span diverse retrieval challenges. Retrieval performance is assessed using mAP@1000, nDCG@10, and Recall@1k, capturing the trade-off between early precision, global ranking quality, and coverage of relevant documents. 

\noindent \textbf{Baselines.} We benchmark our framework against a set of SOTA query reformulation methods spanning traditional, feedback-based, and neural approaches. 
We include traditional pseudo-relevance feedback via RM3~\cite{abdul2004umass,lin2019neural,yang2019critically} and Rocchio~\cite{rocchio1971relevance}, as well as neural reformulation baselines comprising \texttt{GenQR}~\cite{wang2023generative}, an LLM-based keyword-oriented generative model, and \texttt{GenQREnsemble}~\cite{dhole2024genqrensemble}, a prompt-ensemble variant designed to improve keyword coverage.
In addition, we evaluate the integration of our pattern-guided framework, \texttt{ReFormeR}, with state-of-the-art context-based expansion and generation methods, including \texttt{QA-Expand}~\cite{seo2025qa}, which generates relevant questions and pseudo-passages to enrich query context; \texttt{Query2Doc}~\cite{wang2023query2doc} in zero-shot (ZS), few-shot (FS), and chain-of-thought (CoT) configurations, which produces a pseudo-passage that captures the semantics of the query; and \texttt{MUGI}~\cite{zhang2024exploring}, a dense generation-augmented retriever.

\noindent \textbf{Implementation Details.} Retrieval was carried out using BM25 implemented by Pyserini~\cite{pyserini}, selected for its robust integration with large-scale document collections. 
All query reformulation methods and experiments were conducted using the \texttt{Qwen2.5-7B-Instruct} model~\cite{team2024qwen2}, running inference via the vLLM framework \cite{kwon2023efficient}. For \texttt{ReFormeR} implementation, the generation parameters were set to a maximum of 512 tokens and a decoding temperature of 1.0. Baseline results were reproduced following publicly available implementations and settings specified in prior work and where applicable, hyperparameters were adopted directly from published baselines. 

\noindent \textbf{Extracting Reformulation Patterns.} To derive effective reformulation patterns, we employ the \textit{Diamond} subset of the \textit{Matches Made in Heaven} collection released by \cite{arabzadeh2021matches}.
This dataset comprises MS MARCO \cite{nguyen2016ms} train queries paired with their reformulated counterparts that achieve perfect retrieval effectiveness (MRR=1), thereby providing an ideal setting for analyzing the semantic transformations underlying successful query reformulations.
For this purpose, we utilize  \texttt{Qwen2.5-72B} as the underlying LLM for pattern induction.
A total of 10,000 query–reformulation pairs were sampled from the dataset, and the LLM was prompted to traverse these pairs, infer the underlying patterns that explain the improvement of each reformulation, and consolidate recurring rationales into a unified reformulation pattern library using the prompt shown in Figure~\ref{fig:reformulation_prompt}. A summary of final list of consolidated patterns is available in Table~\ref{reason_categories}. A more detailed version of this table is available on our GitHub.

\begin{table}[!t]
\centering

\caption{Consolidated list of reformulation patterns extracted by \texttt{ReFormeR}.}
\label{tab:reason_categories}
\setlength{\tabcolsep}{10pt} 
\renewcommand{\arraystretch}{1.3}
\scalebox{0.77}{
\begin{tabular}{llll}
\toprule
Clarify Intent & Clarify Subject & Conceptual Shift & Contextual Expansion \\
Contextual Restriction & Generalization & Location Specification & Purpose Specification \\
Semantic Clarification & Temporal Adjustment & & \\
\bottomrule
\end{tabular}}

\label{reason_categories}
\end{table}

\subsection{Findings}

Table~\ref{tab:results_table} presents the retrieval effectiveness of \texttt{ReFormeR} compared with both keyword-based and context-aware reformulation approaches across TREC DL 2019, TREC DL 2020, and TREC DL Hard. 
For context-aware baselines, \texttt{ReFormeR} is integrated within their generation pipeline by embedding its pattern-guided reformulated queries into the original prompt context.

\noindent \textbf{Comparison with keyword-based reformulation.} \texttt{ReFormeR} delivers consistent and meaningful improvements over both classical feedback and generative baselines.  
Against traditional methods such as \texttt{RM3} and \texttt{Rocchio}, \texttt{ReFormeR} achieves up to 16\% higher nDCG@10 on TREC DL 2020 and about 26-34\% higher mAP@1k on DL Hard, underscoring the advantage of pattern-guided reformulation over purely statistical expansion.  
While keyword-based feedback models tend to amplify frequent but noisy terms, \texttt{ReFormeR} introduces targeted semantic refinements that align better with the underlying retrieval intent based on the identified reformulation patterns.  
This advantage becomes more evident as query difficulty increases on DL Hard, where queries are sparse or ambiguous, pattern-guided reformulations provide the largest gains by improving precision without sacrificing recall.

Compared to keyword-based generative reformulation methods such as \texttt{GenQR} and \texttt{GenQREnsemble}, \texttt{ReFormeR} achieves consistent and substantial improvements, especially on the challenging TREC DL 2020 and DL Hard benchmarks. On TREC DL 2020, it improves nDCG@10 by 26\% relative to \texttt{GenQREnsemble}, showing greater retrieval consistency through more targeted expansions.  
On DL Hard, \texttt{ReFormeR} yields 11\% and 8\% gains in mAP@1k and nDCG@10 over \texttt{GenQR}, demonstrating its strength in handling underspecified queries.  
Rather than generating unconstrained paraphrases or keyword additions, \texttt{ReFormeR} performs guided query reformulation resulting in more effective retrieval outcomes.

\begin{table}[t]
\centering
\vspace{-4em}
\caption{Retrieval effectiveness in terms of mAP@1k, nDCG@10, and Recall@1k for keyword-based and context-based reformulation methods across three TREC datasets. ReFormeR-enhanced methods are highlighted.}
\label{tab:results_table}

\begin{tabular}{lccccccccc}
\toprule
 & \multicolumn{3}{c}{\textbf{DL 2019}} & \multicolumn{3}{c}{\textbf{DL 2020}} & \multicolumn{3}{c}{\textbf{DL Hard}} \\
 \cmidrule(lr){2-4} \cmidrule(lr){5-7} \cmidrule(lr){8-10}
\textbf{Method} & mAP & nDCG & Recall & mAP & nDCG & Recall & mAP & nDCG & Recall \\
\midrule
\texttt{BM25}                         & 0.290 & 0.497 & 0.745 & 0.288 & 0.488 & 0.803 & 0.164 & 0.290 & 0.678 \\
\texttt{BM25 + RM3}                   & 0.334 & 0.515 & 0.795 & 0.302 & 0.492 & 0.829 & 0.154 & 0.264 & 0.699 \\
\texttt{BM25 + Rocchio}               & \textbf{0.347} & 0.528 & \textbf{0.801} & 0.312 & 0.491 & 0.816 & 0.164 & 0.277 & 0.704 \\
\texttt{GenQR}~\cite{wang2023generative}                       & 0.341 & 0.517 & 0.797 & 0.336 & 0.551 & 0.840 & 0.186 & 0.313 & \textbf{0.719} \\
\texttt{GenQREnsemble}~\cite{dhole2024genqrensemble}          & 0.278 & 0.442 & 0.792 & 0.287 & 0.453 & 0.795 & 0.122 & 0.197 & 0.651 \\
\rowcolor{gray!15}
\texttt{ReFormeR}              & 0.318 & \textbf{0.544} & 0.783 & \textbf{0.339} & \textbf{0.572} & \textbf{0.845} & \textbf{0.207} & \textbf{0.337} & 0.718 \\
\bottomrule

\\[-0.5em]

\toprule
 & \multicolumn{3}{c}{\textbf{DL 2019}} & \multicolumn{3}{c}{\textbf{DL 2020}} & \multicolumn{3}{c}{\textbf{DL Hard}} \\
 \cmidrule(lr){2-4} \cmidrule(lr){5-7} \cmidrule(lr){8-10}
\textbf{Method} & mAP & nDCG & Recall & mAP & nDCG & Recall & mAP & nDCG & Recall \\
\midrule
\texttt{QA-Expand}~\cite{seo2025qa}             & 0.363 & 0.587 & 0.816 & 0.376 & 0.567 & 0.847 & 0.189 & 0.284 & 0.725 \\
\rowcolor{gray!15}
\phantom{MUGI}\,+ \texttt{ReFormeR}             & \textbf{0.406} & \textbf{0.605} & \textbf{0.836} & \textbf{0.382} & \textbf{0.579} & \textbf{0.880} & \textbf{0.209} & \textbf{0.302} & \textbf{0.762} \\
\texttt{Query2Doc (ZS)}~\cite{wang2023query2doc}        & 0.426 & 0.632 & 0.861 & 0.379 & 0.579 & 0.881 & 0.219 & 0.348 & 0.783 \\
\rowcolor{gray!15}
\phantom{MUGI}\,+ \texttt{ReFormeR}             & \textbf{0.438} & \textbf{0.651} & \textbf{0.865} & \textbf{0.410} & \textbf{0.618} & \textbf{0.883} & \textbf{0.227} & 0.342 & 0.756 \\
\texttt{Query2Doc (FS)}~\cite{wang2023query2doc}        & 0.409 & 0.596 & 0.841 & 0.394 & 0.604 & 0.876 & 0.215 & 0.334 & 0.782 \\
\rowcolor{gray!15}
\phantom{MUGI}\,+ \texttt{ReFormeR}             & \textbf{0.449} & \textbf{0.677} & \textbf{0.886} & \textbf{0.413} & \textbf{0.626} & \textbf{0.890} & \textbf{0.233} & \textbf{0.365} & \textbf{0.799} \\
\texttt{Query2Doc (CoT)}~\cite{wang2023query2doc}       & 0.391 & 0.584 & 0.862 & 0.372 & 0.595 & 0.875 & 0.210 & 0.332 & 0.751 \\
\rowcolor{gray!15}
\phantom{MUGI}\,+ \texttt{ReFormeR}             & \textbf{0.428} & \textbf{0.647} & 0.859 & \textbf{0.413} & \textbf{0.611} & \textbf{0.885} & \textbf{0.218} & \textbf{0.353} & \textbf{0.798} \\
\texttt{MUGI}~\cite{zhang2024exploring}         & 0.466 & 0.692 & 0.891 & 0.417 & 0.639 & 0.887 & 0.224 & 0.347 & 0.794 \\
\rowcolor{gray!15}
\phantom{MUGI}\,+ \texttt{ReFormeR}             & 0.466 & 0.684 & 0.887 & \textbf{0.432} & \textbf{0.653} & \textbf{0.901} & \textbf{0.250} & \textbf{0.375} & \textbf{0.814} \\
\bottomrule
\end{tabular}
\vspace{-2em}

\end{table}

\noindent \textbf{Comparison with context-based reformulation.}  
When integrated with advanced context-aware reformulators, \texttt{ReFormeR} consistently boosts retrieval effectiveness across datasets.  
For instance, combining \texttt{QA-Expand} with \texttt{ReFormeR} improves mAP@1k by 12\% on TREC DL 2019 and 11\% on DL Hard, while enhancing recall from 0.847 to 0.880 on TREC DL 2020.  
Similarly, for \texttt{Query2Doc (FS)}, \texttt{ReFormeR} yields up to 13.6\% higher nDCG@10 on TREC DL 2019 and 9\% higher on DL Hard, reflecting stronger early ranking precision.  
The largest benefit emerges under more challenging retrieval settings, where coupling \texttt{MUGI} with \texttt{ReFormeR} achieves an mAP@1k of 0.250 and nDCG@10 of 0.375 on DL Hard. These correspond to relative gains of 11.6\% and 8.1\% over the baseline, establishing new best results among context-based models.
These improvements reveal that pattern-guided reformulation is effective when contextual generation alone is insufficient. While context-aware models rely on document content, \texttt{ReFormeR} injects reformulation patterns that preserve query intent, prevent semantic drift, and enhance interpretability across diverse retrieval conditions.

\section{Concluding Remarks}

This paper presented \texttt{ReFormeR}, a pattern-guided approach that induces a compact, reusable library of query reformulation patterns and applies them to generate targeted query reformulations. By making the reformulation policy explicit through patterns, \texttt{ReFormeR}  delivers consistent gains over feedback and LLM-based reformulation baselines on DL’19, DL’20, and DL-Hard. It also complements context-aware reformulation baseline methods, yielding further improvements when integrated with state-of-the-art methods such as \texttt{QA-Expand}, \texttt{Query2Doc}, and \texttt{MUGI}.

\bibliographystyle{splncs04}

\end{document}